\documentclass[aps,pra,twocolumn]{revtex4-1}
\usepackage{amsmath,amsfonts,amssymb,epsfig}

\begin{document}

\title{Maximally entangled gapped ground state of lattice fermions}
\author{David L.\ Feder}
\email{dfeder@ucalgary.ca}
\affiliation{Institute for Quantum Information Science,
University of Calgary, Alberta T2N 1N4, Canada}

\date{\today}

\begin{abstract}
Entanglement between the constituents of a quantum system is an essential 
resource in the implementation of many quantum processes and algorithms. 
Indeed, universal quantum computation is possible by measuring individual 
qubits comprising highly entangled `cluster states.' In this work it is shown 
that the unique gapped ground state of non-interacting fermions hopping on a 
specially prepared lattice is equivalent to a cluster state, where the 
entanglement between qubits results solely by fermionic indistinguishability
and antisymmetry. 
A deterministic strategy for universal measurement-based quantum computation
with this resource is described. 
Because most matter is composed of fermions, these results suggest that 
resources for quantum information processing might be generic in Nature.
\end{abstract}

\maketitle

\section{Introduction}

In measurement-based quantum computation (MBQC)~\cite{Briegel:2009lr}, an
algorithm proceeds entirely by making projective measurements of successive
particles (qubits or qudits) comprising some highly entangled `resource state.' 
While two-dimensional cluster states are known to be universal resources for 
MBQC~\cite{Raussendorf:2003}, it has been proven that they cannot be the 
unique ground states of any qubit Hamiltonian involving at most two-body 
interactions~\cite{nest:012301}. Because two-body interactions are generic in 
Nature, this limitation has driven a search for other universal resource 
states that might exist as the ground state of a physical system. An 
alternative representation of MBQC using Matrix Product States
(MPS)~\cite{fannes:1992} in one dimension and its extension to higher 
dimensions using Projected Entangled Pair States 
(PEPS)~\cite{verstraete:2004,perez:2007} indicates that there are many states
with much less entanglement than the cluster states that are nevertheless
universal resources~\cite{gross:052315,cai:050503,mora:2010}. Unfortunately,
the fraction of useful resource states is known to be exponentially small in
the number of physical qubits~\cite{gross:190501,bremner:190502}. Worse, it is
presently unknown if any of these states is the unique, gapped ground state of
any physically realizable Hamiltonian. By relaxing various assumptions, it is 
possible to define accessible Hamiltonians whose ground states are universal 
resources for MBQC. One example is a system of spin-5/2 qudits on the 
hexagonal lattice with two-body interactions, where the ground state is unique 
and gapped~\cite{chen:2009}. Another is the AKLT model of an 
antiferromagnet\cite{AKLT} with spin-3/2 qudits on the hexagonal lattice, 
which is gapped but degenerate~\cite{miyake:2011,wei:2011}.

An unrelated strategy for MBQC was devised for indistinguishable free fermions 
using beam splitters and charge detectors~\cite{beenakker:2004} in analogy 
with non-deterministic linear optical quantum computation~\cite{knill:2001}. In 
fact, cluster states can be dynamically generated using this 
approach~\cite{zhang:2006}. The relationship between free fermions and 
entanglement has subsequently been the subject of 
extensive investigations~\cite{amico:2008,eisert:2010}. More recently, the 
relationships between fermions and quantum information are being 
explored~\cite{barthel:2009,jozsa:2010,kraus:2010,pizorn:2010}.
In particular, it has been shown that quantum circuits composed only of 
matchgates~\cite{Valiant:2002,Ramelow:2010} (which are closely related to 
non-interacting fermions) can generate entanglement yet are efficiently 
classically simulatable unless supplemented by an apparently trivial resource, 
the unentangling SWAP gate~\cite{jozsa:2010}.

In this work, the ideas above are combined to show that the non-degenerate and
gapped ground state of a non-interacting fermionic lattice Hamiltonian is 
equivalent to a cluster state comprised of qubits, under the conditions that 
any site has at most one fermion and the lattice is exactly half-filled. The 
qubit registers then correspond to the occupation of lattice sites. The 
entanglement is a direct consequence of fermionic indistinguishability and 
antisymmetry. Measurements correspond to detecting the presence or absence of 
a fermion in a lattice site, equivalent to a restriction to the computational 
basis. In order to simulate a universal quantum computer, one must rotate the 
measurement from the computational $Z$ basis into the $X$ or $Y$ basis via the 
application of a Hadamard gate $H$ or a $\sqrt{X}$ operation (here 
$X$, $Y$, and $Z$ represent the Pauli matrices). Surprisingly, in the 
fermionic model these gates cannot be directly implemented because of the 
intrinsic entanglement; rather, an additional lattice must be introduced that 
allows two fermions encoding different qubits to interact, thereby cancelling 
the effects of the fermionic antisymmetry. Fermionic systems therefore invert 
the usual quantum computing paradigm, in that entanglement is intrinsic but 
general single-qubit operations require particle interactions.

The manuscript is organized as follows. A review of cluster states and the 
measurement-based model for universal quantum computation using them is briefly
reviewed in Sec.~\ref{sec:MBQC}. Sec.~\ref{sec:fermions} introduces the main 
fermionic model, and
demonstrates that in principle the ground state is a universal resource for
MBQC. The implementation of a universal set of quantum gates by measurements
is discussed in Sec.~\ref{sec:practical}, where it is shown that interactions
and additional potentials are required. The implications of the results are 
summarized in Sec.~\ref{sec:conclusions}.

\section{\bf Review of cluster-state quantum computation}
\label{sec:MBQC}

The stabilizers of a given state $|\psi\rangle$ are the operators $S_i$ for 
which one satisfies $S_i|\psi\rangle=|\psi\rangle$ for each $i$. If 
$|\psi\rangle$ is a stabilizer state, then its Hilbert space dimension exactly 
coincides with the number of stabilizers $S_i$. A useful class of stabilizer
states are known as graph states, where qubits are represented by vertices and 
maximal two-qubit entanglement between two qubits is represented by an edge 
connecting the vertices. For a generic graph state $|g\rangle$ on $N$ qubits, 
the $N$ generators for the $2^N$-dimensional group of stabilizers are given by 
the separable operators $S_i=X_i\bigotimes_{j\in{\mathcal N}(i)}Z_j$, where 
${\mathcal N}(i)$ signifies the neighbourhood of site (qubit) $i$, and where 
$X_i$, $Y_i$, and $Z_i$ denote Pauli matrices acting on qubit $i$. A 
Hamiltonian that is guaranteed to have $|g\rangle$ as the lowest energy state 
(with eigenvalue $-N\tau$) is therefore 
\begin{equation}
\hat{H}=-\tau\sum_{i=1}^{N}S_i=-\tau\sum_{i=1}^NX_i\bigotimes_{j\in{\mathcal N}(i)}Z_j.
\label{Hamstab}
\end{equation}
For cluster states, in which the qubits are arranged on a regular lattice, the
coordination number is two and four for one-dimensional (1D) and 
two-dimensional (2D) lattices, respectively. This implies that the 
Hamiltonian above consists of three-spin and five-spin operators in 1D and 2D,
respectively. In fact, this number has been proven to be a minimum, i.e.\ 
there is no Hamiltonian having all terms with fewer spin operations that can 
have $|g\rangle$ as a (non-degenerate) ground state~\cite{nest:012301}.

In the one-dimensional case, the cluster ground state is
\begin{equation}
|C_N\rangle\equiv\prod_{i=1}^{N-1}CZ_{i,i+1}|+\rangle^{\otimes N},
\label{eq:cluster}
\end{equation}
where $CZ_{i,j}\equiv{\rm diag}\left(1,1,1,-1\right)$ is the maximally
entangling controlled-phase gate and $|+\rangle=(|0\rangle+|1\rangle)/\sqrt{2}$
is the $+1$ eigenstate of the $X$ operator. In this representation, it is 
evident that the cluster state is highly entangled: each qubit is maximally 
entangled with its neighbor. The excited states correspond to
$\prod_{j=1}^NZ_j^{k_j}|C_N\rangle$, where $k_j\in\{0,1\}$, and have energies 
$(-N+\ell)\tau$ with $\ell\in\{1,\ldots,2N\}$. The energy gap to the first
excited state is therefore $\tau$, constant in the thermodynamic limit.

To perform universal quantum computation, it suffices to measure qubits in
successive columns of the 2D lattice in the $HR_Z(\theta)$ basis, where
$H$ and $R_Z(\theta)=\exp\left(-i\frac{\theta}{2}Z\right)$ are the 
Hadamard operator and rotation by an angle $\theta$ around the $Z$-axis, 
respectively:
$$H=\frac{1}{\sqrt{2}}\begin{pmatrix}
1 & 1\\
1 & -1\\
\end{pmatrix}
;\;
R_Z(\theta)=\begin{pmatrix}
e^{-i\theta/2} & 0 \\
0 & e^{i\theta/2} \\
\end{pmatrix}
$$
This corresponds to measurements in a basis spanned by the vectors
$|+_{\theta}\rangle=\frac{1}{\sqrt{2}}\left(1+e^{-i\theta}\right)$ and
$|-_{\theta}\rangle=\frac{1}{\sqrt{2}}\left(1-e^{-i\theta}\right)$; 
alternatively, the operation $HR_Z(\theta)$ is first applied to the
qubit, and then it is measured in the computational basis. In both cases, the
result of measuring qubit 1 of a linear cluster state with $N$ qubits is 
$|m\rangle_1\otimes\left[X^mHR_Z(\theta)\right]_2
|+^{\otimes(N-1)}\rangle_{2,3,\ldots N}$, where $m=\{0,1\}$ is the measurement 
outcome and the subscript is the qubit index. The presence of the $X$ gate
reflects the probabilistic nature of the measurement outcome. Three successive
measurements with three independent angles (adapted to previous measurements
because of the possible presence of an $X$ gate) suffices to teleport an 
arbitrary single-qubit gate in 1D cluster states; a universal set of gates 
results in 2D cluster states~\cite{Briegel:2009lr}. \\

\section{Entanglement in Lattice Fermions}
\label{sec:fermions}

\subsection{Fermionic model}
\label{subsec:model}

The fermionic model employed in this work is inspired by experiments
with ultracold atomic gases. Numerous groups have confined both bosonic and
fermionic atoms in periodic potentials formed from interfering lasers, known
as optical lattices~\cite{Bloch:2005kx,morsch:179}. These lattices can confine
different spin components of the same bosonic species in independent,
overlapping optical lattices~\cite{Mandel:2003fk,Mandel:2003}. The structure of
these lattices is highly flexible~\cite{Jaksch:2005}, and periodic double-well
potentials for bosons have been been the subject of particularly intense
investigations~\cite{lee:020402,Anderlini:2007lr,chiara:052333,Folling:2007lr,S.Trotzky01182008}. In another recent development, an ultracold gas consisting of
a mixture of three different spin components of the fermionic atom $^6$Li has
been realized in an optical trap~\cite{ottenstein:203202,Williams:2009}.
Ultracold fermionic gases of both $^{40}$K and $^6$Li have been loaded into
periodic optical lattices with highly controllable geometries and trapping
parameters~\cite{Modugno:2003uq,Kohl:2005qy,Jordens:2008yq,Chin:2006fj}.
It is reasonable to assume that these various techniques could be combined in
the future, with multicomponent fermions in three overlapping double-well
optical lattices, each confining a different spin component.

\begin{figure}
\begin{center}
\epsfig{file=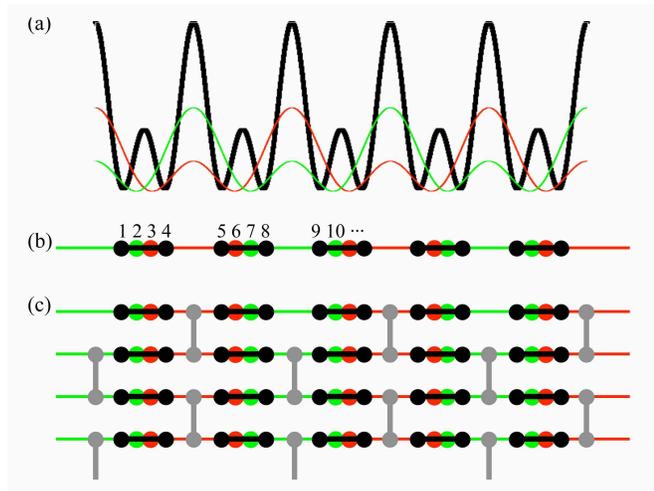,width=\columnwidth,angle=0}
\end{center}
\caption{(Color online) Depiction of the fermionic leapfrog model. Each spin 
projection experiences a unique spatially-dependent potential energy, shown as 
the black, red (dark gray), and green (light gray) curves in (a). Assuming that
the higher potential barriers are effectively impenetrable, one obtains the 
hopping model in (b), with three overlapping double-well sublattices. A 
two-dimensional extension of the one-dimensional model is shown in (c).}
\label{pots}
\end{figure}

Consider the potentials depicted in Fig.~1(a), 
corresponding to three overlapping one-dimensional two-site lattices. It is 
assumed that each 
sublattice can confine only one particular spin projection of a fermion with 
pseudospin $s=1$ (corresponding to real spin $s\geq 3/2$ with all but three
spin projections frozen out); the spin projections will be labeled $\uparrow$,
$o$, and $\downarrow$. The potential shown as a black curve is assumed to 
confine only the spin $o$, and could be obtained using two potentials with 
amplitudes $V_{1,o}$ and $V_{2,o}$ and with one wavelength $\lambda$ half as 
large as another:
$$V_o(x)=V_{1,o}\cos^2\left(\frac{2\pi x}{\lambda}\right)
+V_{2,o}\cos^2\left(\frac{\pi x}{\lambda}\right).$$
Likewise, the potentials seen by the $\uparrow$ and $\downarrow$ spin
projections (green and red curves in Fig.~\ref{pots}(a), respectively)
would be generated by wavelengths that were twice as long as for projection $o$,
and offset relative to each other:
\begin{eqnarray}
V_{\uparrow}(x)&=&V_{1,\uparrow}\cos^2\left(\frac{\pi(x-\lambda)}{\lambda}
\right)+V_{2,\uparrow}\cos^2\left(\frac{\pi(x-\lambda)}{2\lambda}\right);
\nonumber \\
V_{\downarrow}(x)&=&V_{1,\downarrow}\cos^2\left(\frac{\pi x}{\lambda}\right)
+V_{2,\downarrow}\cos^2\left(\frac{\pi x}{2\lambda}\right).\nonumber
\end{eqnarray}
The relative height of the barriers is then set by the ratio 
$V_{1,\sigma}/V_{2,\sigma}$. Within the tight-binding approximation, assuming 
the atoms can only tunnel through the lower barrier of a given sublattice, the 
Hamiltonian becomes
\begin{eqnarray}
\hat{H}&=&-\sum_{j=0}^{N-1}\left(
\tau_{\uparrow}f_{8j+2,\uparrow}^{\dag}f_{8j-1,\uparrow}
+\tau_{o}f_{4j+4,o}^{\dag}f_{4j+1,o}\right. \nonumber \\
&&\qquad\left. +\tau_{\downarrow}f_{8j+6,\downarrow}^{\dag}f_{8j+3,\downarrow}
+{\rm H.c.}\right),
\label{Hamalt0}
\end{eqnarray}
where $f^{\dag}_{i,\sigma}$ ($f_{i,\sigma}$) creates (annihilates) a fermion
with spin projection $\sigma$ at site $i$. The $\tau_{\sigma}$ correspond to 
hopping amplitudes, the exact values of which are unimportant for the present 
work. 

There is no tunneling permitted between different sublattices, but nevertheless 
fermions with different spin projections must anticommute if the sublattices
overlap: if a fermion with one spin projection tunnels past a fermion with 
another spin projection, the total wavefunction picks up a negative sign. This 
is because it is impossible to distinguish if the fermions interchanged their 
spin labels (switching places) during the transit, or remained the same. At 
the same time, it is unnecessary to impose antisymmetrization on fermions that 
can never overlap in practice. For example, a $o$-projection fermion in site 1 
in Fig.~1(b) needs to be antisymmetrized only with fermions in sites 
2, 3, and 4; the tunneling barrier between sites 4 and 5 is assumed to be so
large as to prevent any tunneling for $o$-spin projection atoms.

The relative strengths of the tunneling amplitudes are not important for the 
ground state, as long as they are all non-zero. One may assume that the system
is given sufficiently long to equilibrate compared to the tunneling rates.
Thus one may choose $\tau_o=\tau_{\uparrow}=\tau_{\downarrow}\equiv\tau$ 
without loss of generality. The Hamiltonian~(\ref{Hamalt0}) can be further 
simplified by noting that fermions with a given spin projection are 
constrained to their respective sublattices. The spin-projection index can 
therefore be suppressed, corresponding to spinless (fully polarized) fermions. 
One then obtains the equivalent `leapfrog' Hamiltonian
\begin{equation}
\hat{H}=-\tau\sum_{j=0}^{N-1}\left(f_{2j+4}^{\dag}f_{2j+1}+f_{2j+1}^{\dag}f_{2j+4}
\right),
\label{Hamalt}
\end{equation}
which is the central model of the current work, and is depicted graphically in
Fig.~1(b). Though the Hamiltonian appears to involve third-nearest 
neighbor hopping, it is in fact obtained by nearest-neighbor hopping on an 
array of three overlapping two-site sublattices. 

\subsection{Cluster ground state}
\label{subsec:fermicluster}

Consider only the case where each spinless sublattice is exactly half-filled: 
each connected double-well indexed by $j$ contains exactly one fermion, 
$n_{2j+1}+n_{2j+4}\equiv 1$. Mathematically, this could be accomplished
by applying a suitable projector to the Hamiltonian~(\ref{Hamalt}). 
In an experiment with ultracold atoms, such a configuration could be obtained 
by detecting and selectively moving or removing single 
atoms~\cite{Sortais:2007,Nelson:2007,Beugnon:2007,Karski:2009,Bakr:2009}.

Making use of the Jordan-Wigner transformation
\begin{eqnarray}
f_j&=&\frac{1}{2}\bigotimes_{k=1}^{j-1}Z_k\otimes\left(X_j+iY_j\right);
\nonumber \\
f_j^{\dag}&=&\frac{1}{2}\bigotimes_{k=1}^{j-1}Z_k\otimes\left(X_j-iY_j\right),
\label{JW}
\end{eqnarray}
the Hamiltonian~(\ref{Hamalt}) may be expressed in the spin representation:
\begin{equation}
\hat{H}=-\frac{\tau}{2}\sum_{j=0}^{N-1}Z_{2j+2}Z_{2j+3}\left(
X_{2j+1}X_{2j+4}+Y_{2j+1}Y_{2j+4}\right),
\label{Hamaltspin}
\end{equation}
which is four-local in marked contrast to that in the fermion representation.
Because of the half-filling restriction, the ground state only involves states
with Hamming weight $N$ and therefore only involves combinations of the
encoded registers
\begin{eqnarray}
|\underline{0}_j\rangle\equiv|1_{2j+1}0_{2j+4}\rangle
&=&f_{2j+1}^{\dag}|{\mathcal O}\rangle;\nonumber \\
|\underline{1}_j\rangle\equiv|0_{2j+1}1_{2j+4}\rangle
&=&f_{2j+4}^{\dag}|{\mathcal O}\rangle,\nonumber
\end{eqnarray}
where the state $|{\mathcal O}\rangle$ represents the particle vacuum. In 
other words, the
$|\underline{0}_j\rangle$ register has the single fermion on the first site of
a given double-well indexed by $j$, while the $|\underline{1}_j\rangle$
register corresponds to the fermion in the second site.

One can introduce an encoded basis where the operators become
$\left(X_{2j+1}X_{2j+4}+Y_{2j+1}Y_{2j+4}\right)/2\equiv\underline{X}_j$,
$I_{2j+1}Z_{2j+4}\equiv\underline{Z}_j$, and
$Z_{2j+1}I_{2j+4}\equiv -\underline{Z}_j$ (recall the half-filling condition),
i.e.\ $\underline{X}_j|\underline{0}_j\rangle=|\underline{1}_j\rangle$ and vice
versa, $\underline{Z}_j|\underline{0}_j\rangle=|\underline{0}_j\rangle$, and
$\underline{Z}_j|\underline{1}_j\rangle=-|\underline{1}_j\rangle$. Using
$Z_{2j+2}=I_{2(j-1)+1}Z_{2(j-1)+4}=\underline{Z}_{j-1}$ and
$Z_{2j+3}=Z_{2(j+1)+1}I_{2(j+1)+4}=-\underline{Z}_{j+1}$, the
Hamiltonian~(\ref{Hamaltspin}) in the encoded basis becomes
\begin{equation}
\hat{H}=\tau\sum_{j=1}^{N-1}\underline{Z}_{j-1}\underline{X}_j
\underline{Z}_{j+1} -\tau\underline{Z}_{N-1}\underline{X}_N,
\label{Hamaltspin2}
\end{equation}
with $\underline{Z}_0=\underline{I}_1$. Applying the unitary transformation
$\underline{Z}_j$ for all $j\in\{1,N-1\}$ performs the maps
$\underline{X}_j\to -\underline{X}_j$, transforming (\ref{Hamaltspin2}) into
exactly the same Hamiltonian as the one-dimensional cluster-state Hamiltonian 
on $N$ qubits, Eq.~(\ref{Hamstab}). The ground state of 
Hamiltonian~(\ref{Hamaltspin2}) is therefore immediately obtained by comparing 
with the usual cluster state, Eq.~(\ref{eq:cluster}):
$$|{\rm g.s.}\rangle=\prod_{j=1}^{N-1}\underline{Z}_{j}\underline{CZ}_{j,j+1}
|\mbox{\boldmath $+$}^{\otimes N}\rangle,$$
where $|\mbox{\boldmath $+$}\rangle\equiv\left(|\underline{0}\rangle
+|\underline{1}\rangle\right)/\sqrt{2}$, and the encoded entangling gate is 
defined as $\underline{CZ}_{j,j+1}\equiv \mbox{diag}\left(1,1,1,-1\right)$ in 
the space spanned by the vectors
$|\underline{0}_j\underline{0}_{j+1}\rangle$, 
$|\underline{0}_j\underline{1}_{j+1}\rangle$, 
$|\underline{1}_j\underline{0}_{j+1}\rangle$, and
$|\underline{1}_j\underline{1}_{j+1}\rangle$. It is clear that the ground state
is equivalent to the one-dimensional cluster state, where each qubit is 
maximally entangled to its neighbor. This entanglement is
a direct consequence of fermionic antisymmetry and indistinguishability. The 
excitations correspond to $\underline{Z}_i$ operations on the ground state, 
which cost energy $\tau$.  The energy gap is therefore $\tau$, independent of 
system size.

\subsection{Example: two qubits} 

Consider the consequences of measuring the 
position of one of the two fermions in a two-qubit fermionic cluster state. 
Suppose that the entire system consisted of sites 1, 3, 4, and 6 in Fig.~1(b). 
In the absence of inert 
fermions in the unpaired sites 2 and 5, these sites can be ignored, and the 
ground state of the Hamiltonian~(\ref{Hamalt}) is maximally entangled:
\begin{eqnarray}
|{\rm g.s.}\rangle
&=&\frac{1}{2}\left(f_1^{\dag}+f_{4}^{\dag}\right)\left(f_3^{\dag}+f_{6}^{\dag}
\right)|{\mathcal O}\rangle_{1,3,4,6}\nonumber \\
&=&\frac{1}{2}\left(|1100\rangle+|1001\rangle
-|0110\rangle+|0011\rangle\right)_{1,3,4,6}\nonumber \\
&=&\underline{Z}_1\underline{CZ}_{12}|\mbox{\boldmath $++$}\rangle.
\label{gsex}
\end{eqnarray}
In the Fermi-occupation notation, the ground state appears to be separable;
however, this ignores the fact that fermions on sites 3 and 4 must anticommute
if there is any chance that these fermions can interchange. Consequently,
defining the fermionic occupation states in the natural ordering by increasing label
$|0110\rangle_{1,3,4,6}=f_3^{\dag}f_4^{\dag}|{\mathcal O}\rangle_{1,3,4,6}$,
a negative sign appears on the second line above. If the fermions on site-pairs
$\{1,4\}$ and $\{3,6\}$ could never overlap (i.e.\ there was an impenetrable
barrier between them), then the state would be truly separable and no explicit
antisymmetrization would be required.

Measurement of the first qubit in encoded state $|\underline{0}_1\rangle$ 
corresponds to the application of the projector 
$\left({\mathcal M}_1\right)_1=|1_10_4\rangle\langle 1_10_4|
=f_1^{\dag}|{\mathcal O}\rangle\langle{\mathcal O}|f_1$, yielding
$$\left({\mathcal M}_1\right)_1|{\rm g.s.}\rangle=\frac{1}{\sqrt{2}}\left[
|1100\rangle+|1001\rangle\right]=|\underline{0}\mbox{\boldmath $+$}\rangle.$$
Likewise, applying the other projector $\left({\mathcal M}_2\right)_1
=|0_11_4\rangle\langle 0_11_4|=f_4^{\dag}|{\mathcal O}\rangle
\langle{\mathcal O}|f_4$ yields the state
\begin{eqnarray}
\left({\mathcal M}_2\right)_1|{\rm g.s.}\rangle&=&\frac{1}{\sqrt{2}}\left[
-|0110\rangle+|0011\rangle\right]\nonumber \\
&=&-\underline{Z}_1|\underline{1}\mbox{\boldmath $+$}\rangle
=-|\underline{1}\mbox{\boldmath $-$}\rangle.\nonumber
\end{eqnarray}
The dependence of the second encoded qubit's output state on the outcome of 
the first encoded qubit's measurement is a direct consequence of the 
entanglement of the original fermionic state.

It is important to note, and easy to verify, that this result doesn't depend 
on how one labels the sites. If the two double-sites were instead labeled 
$\{1,3\}$ and $\{4,6\}$, so that no negative sign appeared in the ground 
state~(\ref{gsex}), the same states would still be obtained after measurement. 
This is because the result of the measurement operation on the double-site 
$\{1,3\}$ explicitly depends on the possible existence of a fermion at the 
intervening site 4. If instead the two double-wells were disconnected
(non-overlapping), then there would be no intervening site with which to 
anticommute, and therefore no dependence of the output on the measurement 
outcome. In short, the entanglement arises from the fundamental requirement 
that fermions on the sites 3 and 4 must anticommute when there is the potential 
for these wavefunctions to overlap.

\section{Universal measurement-based quantum computation with lattice fermions}
\label{sec:practical}

\subsection{Single-qubit operations with fermions}

In order to perform universal gate teleportation using the encoded
cluster state, Eq.~(10) in the manuscript, it is important to derive 
expressions for the encoded 
$\underline{X}$, $\underline{Y}$, and $\underline{Z}$-rotations, Hadamard 
operator $\underline{H}$, and measurement operators that one would 
need to apply in order to teleport gates with fermions. 
The encoded $\underline{Z}$-rotation is straightforward:
\begin{eqnarray}
\underline{R}_{Z}(\theta)_j&=&\exp\left(-i\underline{Z}_j\frac{\theta}{2}\right)
=\exp\left(-iZ_{2j+4}\frac{\theta}{2}\right)\nonumber \\
&=&\cos\left(\frac{\theta}{2}\right)-i\sin\left(\frac{\theta}{2}\right)
\left(1-2n_{2j+4}\right),\nonumber
\end{eqnarray}
where $1-2n_{2j+4}=\underline{Z}_j$ is the encoded Z operator.
Because $n_{2j+4}=1-n_{2j+1}$ an equivalent expression would be
$$\underline{R}_Z(\theta)_j=\cos\left(\frac{\theta}{2}\right)
+i\sin\left(\frac{\theta}{2}\right)\left(1-2n_{2j+1}\right).$$
Thus, $\underline{R}_Z(\theta)=e^{-i\theta/2}$ if the fermion is in the first
site and $\underline{R}_Z(\theta)=e^{i\theta/2}$ if it is in the second, as
expected from the qubit encoding. To obtain the encoded 
$\underline{X}$-rotation, one can make use of the relations
\begin{eqnarray}
\underline{X}_j&=&\left(1-2n_{2j+2}\right)\left(1-2n_{2j+3}\right)\nonumber \\
&\times&\left(f^{\dag}_{2j+4}f_{2j+1}+f^{\dag}_{2j+1}f_{2j+4}\right),
\label{fermiX}
\end{eqnarray}
and $\underline{X}_j^2=n_{2j+1}\left(1-n_{2j+4}\right)
+n_{2j+4}\left(1-n_{2j+1}\right)=n_{2j+1}+n_{2j+4}=1$ when each two-site 
lattice has exactly one fermion. One then finds
\begin{widetext}
$$\underline{R}_{X}(\theta)_j=\exp\left(-i\underline{X}_j\frac{\theta}{2}\right)
=\cos\left(\frac{\theta}{2}\right)-i\sin\left(\frac{\theta}{2}\right)
\left(1-2n_{2j+2}\right)
\left(1-2n_{2j+3}\right)\left(f^{\dag}_{2j+4}f_{2j+1}+f^{\dag}_{2j+1}f_{2j+4}
\right).$$
One can write $\underline{Y}_j=-i\underline{Z}_j\underline{X}_j
=iZ_{2j+1}\underline{X}_j$, so that the encoded $\underline{Y}$-rotation is
$$\underline{R}_{Y}(\theta)_j=\exp\left(-i\underline{Y}_j\frac{\theta}{2}
\right)=\cos\left(\frac{\theta}{2}\right)+\sin\left(\frac{\theta}{2}\right)
\left(1-2n_{2j+1}\right)\left(1-2n_{2j+2}\right)\left(1-2n_{2j+3}\right)
\left(f^{\dag}_{2j+4}f_{2j+1}+f^{\dag}_{2j+1}f_{2j+4}\right).$$
Finally, the encoded Hadamard gate is 
$\underline{H}_j=\underline{Z}_j\underline{R}_Y(-\pi/2)_j
=-\left(1-2n_{2j+1}\right)\underline{R}_Y(-\pi/2)_j$:
\begin{equation}
\underline{H}_j=\frac{1}{\sqrt{2}}\left[-1+2n_{2j+1}
+\left(1-2n_{2j+2}\right)\left(1-2n_{2j+3}\right)
\left(f^{\dag}_{2j+4}f_{2j+1}+f^{\dag}_{2j+1}f_{2j+4}\right)\right].
\label{Hadamard}
\end{equation}
\end{widetext}

Note that in marked contrast with the qubit representation, in the fermion 
representation the $\underline{X}_j$, $\underline{Y}_j$, and 
$\underline{H}_j$ operators acting on encoded qubit $j$ (involving the 
two sites labeled by $2j+1$ and $2j+4$) are intrinsically non-local, since 
they depend explicitly on the fermion density at the intervening sites 
$2j+2$ and $2j+3$. Nevertheless they are fully separable (i.e.\ not 
entangling), as will be shown explicitly in the example below. One 
might ask why they are nevertheless non-local. Suppose that the 
$\left(1-2n_{2j+2}\right)\left(1-2n_{2j+3}\right)$ term were dropped in
expression~(\ref{fermiX}) for the encoded $\underline{X}_j$ operator. The 
operation would then na\"\i vely correspond to an unencoded $X_j$ operation, 
but only in the absence of fermions between sites $2j+1$ and $2j+4$. If there 
were in fact one fermion occupying either of 
the intervening sites $2j+2$ or $2j+3$, then hopping between $2j+1$ and $2j+4$ 
would introduce undesired negative signs to certain matrix elements. Of 
course, these are exactly the factors that yield the ground-state entanglement 
in the first place! But they are unwanted in the implementation of 
$X$, $Y$, and $H$ operations, and so need to be cancelled in the 
encoded versions $\underline{X}$, $\underline{Y}$, and $\underline{H}$
if necessary. Consider two encoded qubits, for example. The basis is spanned 
by the vectors $|1100\rangle=f_1^{\dag}f_3^{\dag}|{\mathcal O}_4\rangle$,
$|1001\rangle=f_1^{\dag}f_6^{\dag}|{\mathcal O}_4\rangle$,
$|0110\rangle=f_3^{\dag}f_4^{\dag}|{\mathcal O}_4\rangle$, and
$|0011\rangle=f_4^{\dag}f_6^{\dag}|{\mathcal O}_4\rangle$, the encoded 
Hadamard gate~(\ref{Hadamard}) acting on the first encoded qubit is
\begin{eqnarray}
\underline{H}_1&=&\frac{1}{\sqrt{2}}
\begin{pmatrix}
1 & 0 & 1 & 0\cr
0 & 1 & 0 & 1\cr
1 & 0 & -1 & 0\cr
0 & 1 & 0 & -1\cr
\end{pmatrix}\nonumber \\
&=&\frac{1}{\sqrt{2}}\left(\underline{X}+\underline{Z}\right)\otimes
\underline{I}=\underline{H}\otimes\underline{I},
\end{eqnarray}
which is manifestly separable in matrix form even though it appears not to be
in the fermion representation~(\ref{Hadamard}). 

The measurements are made in the local 
fermion basis, so that 
\begin{eqnarray}
\left({\mathcal M}_1\right)_j=|\underline{0}\rangle_j\langle
\underline{0}|&=&|1_{2j+1}0_{2j+4}\rangle\langle 1_{2j+1}0_{2j+4}|\nonumber \\
&=&f_{2j+1}^{\dag}|{\mathcal O}_2\rangle_{2j+1,2j+4}
\langle{\mathcal O}_2|f_{2j+1}\nonumber
\end{eqnarray}
and
\begin{eqnarray}
\left({\mathcal M}_2\right)_j&=&|\underline{1}\rangle_j\langle
\underline{1}|=|0_{2j+1}1_{2j+4}\rangle\langle 0_{2j+1}1_{2j+4}|\nonumber \\
&=&f_{2j+4}^{\dag}|{\mathcal O}_2\rangle_{2j+1,2j+4}\langle{\mathcal O}_2|
f_{2j+4}.\nonumber
\end{eqnarray}
In other words, the qubit projectors correspond simply to observing the site
indexed by $2j+1$ or $2j+4$ to determine if a fermion is present; if so, it is
projected into the encoded state $|\underline{0}\rangle$ or 
$|\underline{1}\rangle$, respectively.

\subsection{Example: two-qubit gate teleportation} 

It is instructive to 
consider how one would implement one gate teleportation on a two-qubit 
fermionic cluster state. Suppose that the entire system consisted of sites 1, 
3, 4, and 6 in Fig.~1(b) as in the two-qubit example in the previous section,
with ground state given by Eq.~(\ref{gsex}). Suppose that all sources of 
energy exchange between the system and the
environment are eliminated, other than those associated with applied rotations
and measurements, so that $|\psi\rangle\equiv|{\rm g.s.}\rangle$. In 
order to teleport an arbitrary single-qubit gate in a ordinary 1D cluster 
state, one needs to perform a series of measurements in the 
$HR_Z(\theta)$ basis. The first qubit is to be rotated first by the encoded 
gate $\underline{R}_Z(\theta)_1$, then by $\underline{H}_1$, and then measured 
in the computational basis spanned by the states
$|\underline{0}\rangle=f_1^{\dag}|{\mathcal O}\rangle_{1,4}$ and
$|\underline{1}\rangle=f_4^{\dag}|{\mathcal O}\rangle_{1,4}$. Evidently the
applied rotations on the first qubit will push $|\psi\rangle$ away from the
ground state, though the fermions in all the other double wells remain in 
eigenstates of the original Hamiltonian. As a consequence, the state of the 
first qubit will oscillate at a frequency $\tau_{14}/\hbar$. All states and
operations can therefore be considered only at $2\pi/\tau_{14}$ intervals.
This assumption is not essential, however, as discussed below in 
Sec.~\ref{subsec:practical}.

A $\underline{Z}_1$-rotation on the first encoded qubit yields
\begin{eqnarray}
|\psi'\rangle&=&\underline{R}_Z(\theta)|\psi\rangle=\frac{1}{2}\left(f_1^{\dag}
+e^{i\theta}f_{4}^{\dag}\right)\left(f_3^{\dag}+f_{6}^{\dag}\right)
|{\mathcal O}\rangle_{1,3,4,6}\nonumber \\
&=&\frac{1}{2}\left[|1100\rangle+|1001\rangle+e^{i\theta}\left(-|0110\rangle
+|0011\rangle\right)\right]_{1,3,4,6},\nonumber
\end{eqnarray}
ignoring an overall phase $e^{-i\theta/2}$; subsequent application of an 
encoded Hadamard operation gives
\begin{eqnarray}
|\psi''\rangle&=&\underline{H}_1|\psi'\rangle=\frac{1}{2\sqrt{2}}\left[
\left(1-e^{i\theta}\right)\left(|1100\rangle+|0011\rangle\right)\right. 
\nonumber \\
&+&\left.\left(1+e^{i\theta}\right)\left(|1001\rangle+|0110\rangle\right)
\right]_{1,3,4,6}.\nonumber
\end{eqnarray}
Measurement of the first qubit in encoded state $|\underline{0}_1\rangle$ 
corresponds to the application of the projector 
$\left({\mathcal M}_1\right)_1=|1_10_4\rangle\langle 1_10_4|
=f_1^{\dag}|{\mathcal O}\rangle\langle{\mathcal O}|f_1$:
\begin{eqnarray}
\left({\mathcal M}_1\right)_1|\psi''\rangle&=&\frac{1}{2}\left[
\left(1-e^{i\theta}\right)|1100\rangle
+\left(1+e^{i\theta}\right)|1001\rangle\right]\nonumber \\
&=&\frac{1}{2}|\underline{0}_1\rangle\left[\left(1-e^{i\theta}\right)
|\underline{0}_2\rangle+\left(1+e^{i\theta}\right)|\underline{1}_2\rangle
\right]\nonumber \\
&=&\underline{H}_2\underline{Z}_2\underline{R}_Z(\theta)_2
|\underline{0}\mbox{\boldmath $+$}\rangle.\nonumber
\end{eqnarray}
Likewise, applying the other projector $\left({\mathcal M}_2\right)_1
=|0_11_4\rangle\langle 0_11_4|=f_4^{\dag}|{\mathcal O}\rangle
\langle{\mathcal O}|f_4$ yields the state
\begin{eqnarray}
\left({\mathcal M}_2\right)_1|\psi''\rangle&=&\frac{1}{2}\left[
\left(1+e^{i\theta}\right)|0110\rangle
+\left(1-e^{i\theta}\right)|0011\rangle\right]\nonumber \\
&=&\frac{1}{2}|\underline{1}_1\rangle\left[\left(1+e^{i\theta}\right)
|\underline{0}_2\rangle+\left(1-e^{i\theta}\right)|\underline{1}_2\rangle
\right]\nonumber \\
&=&\underline{X}_2\underline{H}_2\underline{Z}_2
\underline{R}_Z(\theta)_2|\underline{1}\mbox{\boldmath $+$}\rangle.\nonumber
\end{eqnarray}
Thus, the gate teleportation using fermions proceeds exactly as in the usual
spin-qubit case, though with an additional overall $\underline{Z}$ operator
reflecting the original ground state~(\ref{gsex}).

\subsection{Two-dimensional cluster state}
Though the Jordan-Wigner transformation is not uniquely defined in two 
dimensions, the structure of the qubit encoding carries naturally over to
dimensions higher than one. A two-dimensional version of the leapfrog 
Hamiltonian is shown in Fig.~\ref{pots}(c). Two interpenetrating two-site 
lattices (shown in grey) that are staggered relative to each other are 
superimposed on an array of one-dimensional leapfrog chains, similar to those 
discussed above. Each double-well of the new lattices serves to connect 
adjacent one-dimensional chains, so the sites are situated between a given 
connected pair in the chains. Again, only a single fermion is allowed to 
occupy any given pair of sites of the new lattices. The labelling of the sites 
can be chosen in any convenient way. When writing out the full ground state in 
terms of the fermion field operators, one need only keep in mind which sites 
interleave others. 

\begin{figure}
\begin{center}
\epsfig{file=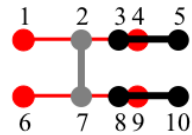,width=0.3\columnwidth,angle=0}
\end{center}
\caption{(Color online) A subgraph of the two-dimensional leapfrog model, with
the two overlapping sublattices corresponding to horizontal red (gray) and
black links.}
\label{pots2D}
\end{figure}

In order to demonstrate that the ground state of the fermionic leapfrog 
Hamiltonian is computationally universal, it remains to be shown that the
presence of the vertical links has the effect of entangling the two horizontal
chains. Consider for this purpose a two-dimensional subgraph of the full 
lattice configuration shown in Fig.~\ref{pots}(c), in which two encoded 
two-qubit cluster states are coupled together by a single vertical 
link as shown in Fig.~\ref{pots2D}. The sites on the top row are labeled 1 
through 5 with site 2 the upper partner of the vertical double well, and 
the bottom three sites are labeled 5 through 10 with site 7 the lower partner 
of site 2. The Hamiltonian corresponding to this subgraph is
$$H_{\rm 2D}=-\tau\left(f_4^{\dag}f_1+f_5^{\dag}f_3+f_7^{\dag}f_2
+f_9^{\dag}f_6+f_{10}^{\dag}f_8+{\rm H.c.}\right).$$
It is simple to verify that the ground state is
\begin{eqnarray}
|{\rm g.s.}\rangle_{\rm 2D}&=&\frac{1}{\sqrt{32}}
\left(f_1^{\dag}+f_4^{\dag}\right)
\left(f_3^{\dag}+f_5^{\dag}\right)
\left(f_2^{\dag}+f_7^{\dag}\right)\nonumber \\
&\times&\left(f_6^{\dag}+f_9^{\dag}\right)
\left(f_8^{\dag}+f_{10}^{\dag}\right)|{\mathcal O}\rangle.\nonumber
\end{eqnarray}

If the vertical link (sites 2 and 7) truly entangles the two horizontal chains,
then measuring the fermions in all the double wells labeled by $\{1,4\}$, 
$\{6,9\}$, and $\{2,7\}$ should yield entanglement between the two encoded 
qubits associated with the double wells labeled by $\{3,5\}$ and $\{8,10\}$, 
even though there is no explicit link remaining between them. The $\{1,4\}$ 
and $\{6,9\}$ sites are measured in the encoded $X$ basis, corresponding to the
application of an encoded Hadamard gate followed by a computational-basis
measurement. The $\{2,7\}$ sites are measured in the encoded $Y$ basis, i.e.\ 
the $\{2,7\}$ sites are first transformed by $\sqrt{\underline{X}}$ and then 
measured in the computational basis. The result doesn't depend on the order in 
which these operations and measurements are made, because only Pauli byproduct
operators are teleported. Ignoring overall phases the output is 
\begin{eqnarray}
|\mbox{out}\rangle&=&|m_1m_2m_3\rangle_{1,2,4,6,7,9}\nonumber \\
&\otimes&\left[
\underline{\sigma}_{12}\underline{H}_1\sqrt{\underline{Z}_1}\underline{H}_2
\sqrt{\underline{Z}_2}\underline{CZ}_{12}
|\mbox{\boldmath $++$}\rangle\right]_{3,5,8,10},\nonumber
\end{eqnarray}
where $m_i=\{0,1\}$ are the measurement outcomes and
$$\underline{\sigma}_{12}=\{\underline{X}_1,
\underline{X}_2,\underline{Z}_2,\underline{X}_1\underline{Y}_2,
\underline{Y}_1\underline{X}_2,\underline{Z}_1,\underline{Z}_1\underline{Y}_2,
\underline{Y}_1\underline{Z}_2\}$$
are local encoded Pauli byproduct operators corresponding to the eight possible
outcomes. Thus, the result of these measurements is to maximally entangle the 
two qubits $|\mbox{\boldmath $+$}\rangle_{3,5}$ and 
$|\mbox{\boldmath $+$}\rangle_{8,10}$. If
the two qubits encoded on sites $\{1,4\}$ and $\{6,9\}$ were originally in the
states $|\underline{\psi}_1\rangle$ and $|\underline{\psi}_2\rangle$, then the
result would be the same as that above except with
$|\mbox{\boldmath $++$}\rangle_{3,5,8,10}\to
|\underline{\psi}_1\underline{\psi}_2\rangle_{3,5,8,10}$.

\subsection{Practical Considerations}
\label{subsec:practical}

The two-qubit example demonstrates how one-dimensional fermionic cluster states 
can be used in order to perform universal gate teleportation. Nevertheless, it 
is not 
apparent how the various encoded operations --- the $\underline{R}_Z(\theta)$, 
$\underline{H}$, $\underline{R}_X(\pi/2)=\sqrt{\underline{X}}$, and 
measurements --- would be implemented in practice. 
Of particular concern is that the encoded Hadamard gate and the $\underline{X}$
and $\underline{Y}$ rotations are explicitly non-local in the fermion 
representation, as shown above [cf.\ Eqs.(\ref{fermiX}) and (\ref{Hadamard})]: 
all require local hopping 
on a given pair of sites in a sublattice with an amplitude dependent on the 
occupancy of intervening qubits on different sublattices. 
As shown below, it is possible to implement all encoded operations but only if
one allows the fermions to explicitly interact.

Consider again the two-qubit example discussed above. All local gates on the
first qubit can be performed by the application of the additional two-site
hopping Hamiltonians
\begin{equation}
\hat{H}_a\left(\tau_{14},V_1,V_4\right)=\tau_{14}\left(f_4^{\dag}f_1
+f_1^{\dag}f_4+V_1n_1+V_4n_4\right)
\label{Ham1}
\end{equation}
and
\begin{eqnarray}
\hat{H}_b\left(\tau_{34},g,V_3,V_4\right) &=& \tau_{34}\left(f_{4,\sigma}^{\dag}
f_{3,\sigma}+f_{3,\sigma}^{\dag}f_{4,\sigma}+gn_{3,\sigma}n_{3,\sigma'}\right.
\nonumber \\
& &\qquad \left. +V_3n_{3,\sigma}+V_4n_{4,\sigma}\right),
\label{Ham2}
\end{eqnarray}
for particular choices of parameters and times. The Hamiltonian~(\ref{Ham1})
is the same as Eq.~(\ref{Hamalt}), with the addition of potential energies 
$V_1$ and $V_4$ on the two sites of the first double-well. The spin indices 
$\sigma$ and $\sigma'$ in Hamiltonian~(\ref{Ham2}) represent the fermionic 
spin projection for encoded qubits 1 and 2, respectively. Thus, the 
fermion encoding qubit 1 is allowed to tunnel between sites 4 and 3 and 
possibly interact with the inert fermion on site 3.

A fermion may or may not be present on site 3 (if not, then it occupies site 
6), so the Hamiltonians~(\ref{Ham1}) and (\ref{Ham2}) act on two independent 
two-dimensional blocks. 
For $H_a$, the states in the first block are
$\left\{f_1^{\dag}f_3^{\dag}|{\mathcal O}\rangle\,,f_3^{\dag}f_4^{\dag}
|{\mathcal O}\rangle\right\}$, while those in the second block are 
$\left\{f_1^{\dag}|{\mathcal O}\rangle\,,f_4^{\dag}|{\mathcal O}\rangle
\right\}$. In the first block, a fermion hopping between sites 1 and 4 must
cross the fermion in site 3, acquiring a negative sign in the process; the
sign change is absent in the second block. But all single-qubit operations 
on the first qubit must be truly local: the resulting gate $\exp(-itH)$ cannot 
depend on the state of the second qubit, i.e.\ the possible presence of a 
fermion on site 3. For $H_b$, the two states in the first block correspond to 
$\left\{f_{3,\sigma'}^{\dag}f_{3,\sigma}^{\dag}|{\mathcal O}\rangle\,,
f_{3,\sigma'}^{\dag}f_{4,\sigma}^{\dag}|{\mathcal O}\rangle\right\}$, 
while those in the second block are $\left\{f_{3,\sigma}^{\dag}
|{\mathcal O}\rangle\,,f_{4,\sigma}^{\dag}|{\mathcal O}\rangle\right\}$. No 
fermions cross in either block, but additional phases can result from the
interaction in site 3 of a fermion with spin $\sigma$ and one with spin 
$\sigma'$.

The most straightforward gate is an arbitrary $Z$-axis rotation on the first 
qubit, which corresponds to turning on $H_a$ for a given time: 
$\underline{R}_Z(-V_4t)_1 =\exp\left[-itH_a\left(0,0,V_4\right)\right]$ for 
either block, ignoring an overall unimportant phase. In practise, a strong
repulsive potential would be applied to the barrier between sites 1 and 4 to
turn off tunneling, and another potential applied to site 4. Next consider the
implementation of the encoded Hadamard gate $\underline{H}_1$. Applying
$H_a\left(\tau_{14},-1+\sqrt{2},1+\sqrt{2}\right)$ for a time 
$t=\pi/\sqrt{8}\tau_{14}$ yields $\underline{H}_1$ in the first block 
and $\underline{Z}_1\underline{H}_1\underline{Z}_1$ in the second.
This gate is equivalent to $\underline{CZ}_{12}\underline{H}_1
\underline{CZ}_{12}$ which is not the desired Hadamard gate.
Likewise, the application of 
$H_a\left(\tau_{14},0,0\right)$ for a time $t=\pi/4\tau_{14}$ yields
$\underline{CZ}_{12}\underline{R}_X(\pi/2)_1\underline{CZ}_{12}$ rather than
$\underline{R}_X(\pi/2)_1$. In order to obtain the desired gates, one can 
apply $H_b$ both before and after the $H_a$ above. Choosing 
$H_b\left(\tau_{34},\frac{4}{\sqrt{3}},-\frac{1}{\sqrt{3}},\frac{1}{\sqrt{3}}
\right)$ for a time $t=\sqrt{3}\pi/2\tau_{34}$ yields $\underline{I}_1$ in the 
first block and $\underline{Z}_1$ in the second, i.e.\ the gate 
$\underline{CZ}_{12}$. All parameters in $H_a$ are assumed to be zero during 
the operation of $H_b$. 
Because $H_b$ also applies a possible phase to the first
register of the second qubit, a repulsive $\sigma'$ potential should be 
applied between sites 4 and 6 during its application and removed thereafter.
The majority of the state thereby remains in the ground state of the primary 
governing Hamiltonian.

It is interesting to note that any attempt to perform a local operation 
$\underline{U}_1$ within the inital non-interacting fermionic model instead
yields $\underline{CZ}_{12}\underline{U}_1\underline{CZ}_{12}$, which is a
two-qubit matchgate modulo local operations. For example, attempting
$\underline{U}_1=\underline{H}_1$ produces a maximally entangling gate 
that is locally unitary-equivalent to the matchgate 
${\rm G}_{\rm HH}$~\cite{Ramelow:2010}. In order to obtain a true local 
operation, required for universal MBQC when restricted to computational basis
measurements, one must include some kind of fermion-fermion interaction. In the 
approach above, an additional hopping Hamiltonian with an on-site interaction 
yields an explicit $\underline{CZ}_{12}$ which can cancel the effective 
$\underline{CZ}_{12}$ arising solely from fermionic statistics. Because one 
can construct the SWAP gate from combinations of Hadamards and $CZ$ gates, 
this approach is the measurement-based analogue of the fermionic circuit model 
found previously~\cite{jozsa:2010}.

\section{Discussion and Conclusions}
\label{sec:conclusions}

Through the `leapfrog' model of fermions hopping across one another in a 
lattice, statistical antisymmetry and entanglement are shown to be equivalent 
quantities in fermionic systems. As a result, the ground state has maximal 
entanglement between all neighboring qubits, which is inherently insensitive 
to decoherence as long as the temperature remains lower than the 
energy gap. In principle, such a resource could be used for a variety of 
distributed quantum information protocols. Yet, universal MBQC using this 
resource is impossible under a realistic model where only computational basis 
measurements are allowed. A universal set of gates can be obtained only when
an additional local Hamiltonian is added that allows the fermions to
explicitly interact through collisions. Note also that any 
double-well sites with zero or two fermions behave as defects (holes) in the 
cluster state; universal quantum computation is still possible as long as 
connectivity of the state exceeds the percolation threshold~\cite{Kieling:2007}.

The results suggest that the ground states of interacting fermionic systems
could generically be resources for quantum information processing within the
MBQC framework. One requires only that the constituent fermionic wavefunctions
have the opportunity to overlap one another, and interact in the process.
Because these criteria arguably apply to almost all physical systems, 
discovering other suitable candidate Hamiltonians might be relatively 
straightforward, though finding convenient encodings for the quantum 
information might still pose a significant challenge. \\
\\

\acknowledgments
I would like to thank Aephraim Steinberg, Jens Eisert, and Gora 
Shlyapnikov for stimulating discussions. I am particularly grateful to the 
hospitality of the Laboratoire de Physique Th\' eorique et Mod\` eles 
Statistiques at the Universit\' e de Paris-Sud where the main ideas were worked
out. This work was supported by the Natural Sciences and Engineering Research 
Council of Canada.



\end{document}